\documentclass[journal,twoside,web]{ieeecolor}
\usepackage{generic}
\usepackage{cite}
\usepackage{amsmath,amssymb,amsfonts}
\usepackage{algorithmic}
\usepackage{graphicx}
\usepackage{algorithm,algorithmic}
\usepackage{hyperref}
\hypersetup{hidelinks=true}
\usepackage{textcomp}

\usepackage{xcolor}

\usepackage{bm}

\def\BibTeX{{\rm B\kern-.05em{\sc i\kern-.025em b}\kern-.08em
		T\kern-.1667em\lower.7ex\hbox{E}\kern-.125emX}}
\begin{document}
	\title{Model Predictive Control for Tracking Bounded References With Arbitrary Dynamics}
	\author{
		Shibo Han, Bonan Hou, Yuhao Zhang, Xiaotong Shi, Xingwei Zhao
		\thanks
		{This work was supported in part by the China Scholarship Council under Grant 202106160015, and the National Natural Science Foundation of China under Grant 52422501. } 
		\thanks
		{Shibo Han is with the Department of Mechanical Engineering, National University of Singapore, 117576 Singapore (e-mail: e0846825@u.nus.edu).}
		\thanks
		{Bonan Hou is with the Engineering Systems and Design, Singapore University of Technology and Design, 487372 Singapore (e-mail: bonan\_hou@sutd.edu.sg).}
		\thanks{Yuhao Zhang and Xingwei Zhao are with the State Key Laboratory of Intelligent Manufacturing Equipment and Technology, Huazhong University of Science and Technology, Wuhan 430074, China (e-mail: yuhao\_zhang@hust.edu.cn, zhaoxingwei@hust.edu.cn).}
		\thanks
		{Xiaotong Shi is with the School of Artificial Intelligence, Hubei University, Wuhan 430062, China (e-mail:xiaotongshi@hubu.edu.cn).}
	}
	
	\maketitle
	
	\begin{abstract}
		In this article, a model predictive control (MPC) method is proposed for constrained linear systems to track bounded references with arbitrary dynamics.
		Besides control inputs to be determined, artificial reference is introduced as additional decision variable, 
			which serves as an intermediate target to cope with sudden changes of reference and enlarges domain of attraction.
		Cost function penalizes both artificial state error and reference error, while terminal constraint is imposed on artificial state error and artificial reference.
		We specify the requirements for terminal constraint and  cost function to guarantee recursive feasibility of the proposed method and asymptotic stability of tracking error.
		Then,  periodic and non-periodic references are considered and the method to determine required cost function and terminal constraint is proposed.
		Finally, the efficiency of the proposed MPC controller is demonstrated with simulation examples.
	\end{abstract}
	
	
	\section{Introduction}		\label{sec:introduction}
	\IEEEPARstart{M}{odel} predictive control (MPC) is one of the most important control methods because of its ability in controlling systems with constraints and minimizing a given performance index \cite{kouvaritakis2016model}.
	At each time step, MPC controller solves a finite-horizon constrained optimization problem to determine the control input. 
	Under mild assumptions, MPC controller guarantees recursive feasibility and asymptotic stability of the closed-loop system \cite{mayne2000constrained}.

	When a controller is proposed to solve tracking problems, sudden changes of reference should be taken into consideration \cite{kohler2024analysis},\cite{limon2008mpc},\cite{limon2015mpc}.
	For example, the reference is supposed to be constant but can change sometime and is actually piecewise constant.
	If sudden changes are not considered, a time-consuming re-formulation of the optimization problem is needed at the switching step.
	Moreover, reference can switch to a distant value, which renders the constrained optimization problem infeasible and interrupts MPC algorithms. 
	For specific types of references, for example, constant references \cite{limon2008mpc},\cite{limon2018nonlinear}, \cite{soloperto2022nonlinear},
	and periodic references \cite{limon2015mpc},\cite{sanchez2023tracking}, \cite{franze2022reference},
	MPC controllers have been proposed to deal with the challenges mentioned above.
	Other types of references, such as parameterized regular curves \cite{faulwasser2015nonlinear}, 
	sinusoidal references \cite{krupa2025harmonic}, and polynomial references \cite{cordero2021development}, have been investigated, as well. 
	Besides these types of references, there are still plenty of references to be considered.
	
	To cope with more general references, there have been attempts at tracking references with arbitrary dynamics. 
	Based on the assumption of recursive feasibility, MPC controller proposed in \cite{maeder2010offset} ensures offset-free tracking of references with arbitrary dynamics.
	To achieve error-bounded tracking, MPC structure is utilized in \cite{yuan2019error} based on robust control invariant set.
	The nonlinear MPC controller proposed in \cite{kohler2021constrained} is extended to track references with arbitrary dynamics when there are no constraints and the prediction horizon is large.
	The MPC controller proposed in \cite{wang2023constrained} utilizes a terminal constraint imposed on terminal cost and ensures the convergence of tracking error.
	However, all the above works do not consider the feasibility of controller when reference changes.
	Another related work is given in \cite{ong2020governor} where a reference-governor-based controller is utilized. 
	When only input constraint is imposed, an additional step is conducted to enlarge its attraction region.
	In summary, the problem of tracking references with arbitrary dynamics still requires further investigation for the practical application of MPC controller.
	
	To this end, an MPC controller is proposed for constrained linear systems to track bounded references with arbitrary dynamics. 
	The proposed MPC controller introduces artificial reference as additional decision variable,
		which makes constraints of optimization problem independent of the given reference.
	Thus, feasibility of the optimization problem will not be destroyed when reference switches at some steps. 
	Then, terminal constraint is imposed on the augmented system of  artificial state error and artificial reference, 
	while cost function is designed to penalize artificial state error and reference error. 
	However, artificial reference introduces difficulties in determine terminal constraint and cost function such that recursive feasibility and asymptotic stability can be guaranteed.
	Thus, both periodic and non-periodic references are analyzed and method to determine the weight matrices of the quadratic cost function and terminal constraint is investigated.
	Contributions of this paper are summarized as follows.
	\begin{itemize}
		\item An MPC algorithm incorporating artificial reference as additional decision variable is proposed for constrained systems to track bounded references with arbitrary dynamics.
			  The proposed algorithm solves a predefined quadratic optimization problem which remains feasible even when reference switches at some steps.
		\item We specify the requirements for terminal constraint and cost function to guarantee recursive feasibility and asymptotic stability when artificial reference is introduced.
			  Meanwhile, methods to determine cost function and finitely determined terminal constraint satisfying those requirements are proposed to facilitate the implementation of the proposed MPC algorithm. 
		\item The proposed method requires fewer decision variables when the period of reference is large, thus reducing online computational burden.
			  Moreover, it can also handle non-periodic references generated by a linear exosystem.
	\end{itemize}

	The rest of this article is organized as follows.
	Section II 		formulates the problem and presents preliminary results.
	Section III  	presents the design of the proposed MPC method and requirements for cost function and terminal constraints.
	Section IV		presents method to determine cost function and terminal constraint satisfying those requirements in both periodic and non-periodic cases.
	Section V 		provides examples to demonstrate the effectiveness of the proposed method.
	Section VI		concludes this article.
	
	\textbf{Notation :}
	The sets of integers, real numbers and complex numbers are denoted as $\mathbb{N}, \mathbb{R}$ and $\mathbb{C}$, respectively. 
	$\mathbb{N}_a^b = \{x \in \mathbb{N} | a \leq x \leq b \}$.
	The sets of positive integers and non negative integers are denoted as $\mathbb{N}^+$ and $\mathbb{N}_0^+$, respectively.
	The conjugate and amplitude of a complex number $\lambda$ are denoted as $\mathring{\lambda}$ and $\left | \lambda \right |$, respectively.
	A block diagonal matrix $S$ with diagonal blocks $S_i, i\in\mathbb{N}_1^{i_0}$ is denoted as $S = diag(S_1, S_2, ..., S_{i_0})$.
	The transpose and conjugate transpose of a matrix $S$ are denoted as $S'$ and $S^H$.
	Eigenvalue and determinant of a square matrix $S$ are denoted as $\lambda(S)$ and $\mathrm{det}(S)$, respectively.
	$\left \| x \right \| _T = \sqrt{x'T x}$, $\left \| x \right \| = \sqrt{x' x}$.
	$T$ is positive definite if $\left \| x \right \| _T^2 >0, \forall x \neq 0$, which is denoted as $T \succ 0$.
	A polytope is a convex set which is expressed as $\mathbb{X} = \{x|H_{ieq}x \leq h_{ieq} \}$. 
	An ellipsoid is notated as $\mathcal{E}_T(r_0) = \{ r | r'Tr \leq \left \| r_0 \right \| _T^2\}$.
	For non-empty set $\mathcal{X}_1$ and $\mathcal{X}_2$, 
	Pontryagin set difference is defined as $\mathcal{X}_1 \ominus \mathcal{X}_2 = \{ x| x \oplus \mathcal{X}_2 \subseteq \mathcal{X}_1 \}$.
	If $\mathcal{X}_2$ contains only one element $x_2$, $\mathcal{X}_1 \ominus \mathcal{X}_2$  simplifies to $\mathcal{X}_1 \ominus x_2$.
	Saturation function $\mathrm{sat}(x)$ is defined as $\mathrm{sat}(x) = \min\{x,1\}$.	
	
	\section{Problem Formulation and Preliminaries}		\label{sec:problemFormulationAndPreliminaries}
	Consider the discrete-time linear system described by
	\begin{align}		\label{eqn:actualSystem}
		x(t+1)=A x(t)+B u(t),
	\end{align}
	where $x(t)\in\mathbb{R}^n,u(t)\in\mathbb{R}^m$ are state and control input, respectively. 
	$A$ and $B$ are matrices with compatible dimensions. 
	Meanwhile, there are some constraints on system (\ref{eqn:actualSystem}), which are expressed as
	\begin{align}		\label{eqn:systemConstraints}
		\begin{bmatrix}x'(t) & u’(t)\end{bmatrix}' \in \mathcal{Z} \subset \mathbb{R}^{n+m}.
	\end{align}
	
	The following assumptions are made on system (\ref{eqn:actualSystem}).
	
	(\textbf{A1}) $x$ is measurable.
	
	(\textbf{A2}) The pair $(A, B)$ is controllable, and $K \in \mathbb{R}^{n \times m}$ ensures that $A_\mathrm{cl}=A+BK$ is Schur.
	
	(\textbf{A3}) $\mathcal{Z}$ is a bounded polytope containing the origin in its interior.
	
	Reference $r(t)$ is generated by a linear exosystem given as 
	\begin{align}		\label{eqn:referenceSystem}
		r(t+1) = S r(t),
	\end{align}
	where $r(t) \in \mathbb{R}^{q}, S \in \mathbb{R}^{q \times q}$.
	Meanwhile, it is possible that  $r(t_i+1) \neq S r(t_i)$ at some steps $t_i \in \mathbb{N}^+$, that is, reference switches to $r(t_i+1)$.
	For physical application, controller should remain feasible in this case.	
	
	Tracking error $e(t) \in \mathbb{R}^p$ is given as
	\begin{align}		\label{eqn:eCxQr}
		e(t) = y(t) - y_r(t),
	\end{align}
	where $y(t) = Cx(t)$ is the output of system (\ref{eqn:actualSystem}), $y_r(t) = Q^e r(t)$ is the desired output.
	$C \in \mathbb{R}^{p \times n}$, $Q^e \in \mathbb{R}^{p \times q}$.
	
	Assumptions about the system to be controlled and the reference are given below.
	
	\textbf{(A4)}	Exosystem (\ref{eqn:referenceSystem}) is Lyapunov stable and $\left | \lambda(S)  \right | = 1$.
	
	\textbf{(A5)}	Rows of 
	$\begin{bmatrix}
		A - \lambda I & B  \\
		C             & 0
	\end{bmatrix}$ are linearly independent for all $\lambda$ where $\lambda$ is the eigenvalue of $S$.
	
		 
		\textbf{Remark 1:}
		As bounded reference is considered, (A4) requires exosystem (\ref{eqn:referenceSystem}) to be Lyapunov stable.
		Further, eigenvalues of $S$ should be on or inside the unit circle.
		Meanwhile, those stable components of reference corresponding to eigenvalues inside the unit circle converge to zero after sufficient long time and will not affect the tracking error.
		Thus, we consider those $S$ whose eigenvalues are all on the unit circle.
		

	The objective of this article is to design a nonlinear control law $\pi \big( x(t), r(t) \big)$ such that	
	(i)  constraint (\ref{eqn:systemConstraints}) is satisfied all the time,	
	(ii) tracking error (\ref{eqn:eCxQr}) converges to zero if the given reference is admissible.
	
	When there are no constraints on system (\ref{eqn:actualSystem}), 
	a linear state feedback controller can be employed to ensure the asymptotic convergence of tracking error,
	which is concluded below.
	
		
		\textbf{Lemma 1:}
		(\cite{knobloch2012topics}, Chapter 1 )
		If assumptions (A1), (A2), (A4), and (A5) hold,
		there exist non-zero matrices $\Pi \in \mathbb{R}^{n \times q}$ and $\Gamma \in \mathbb{R}^{p \times n}$ such that
		\begin{subequations}
			\begin{align}
				A \Pi + B \Gamma &= \Pi S, \\
				C \Pi - Q^e        &= 0.
			\end{align}
		\end{subequations}
		Meanwhile, the state feedback controller 
		\begin{align} \label{eqn:IMPcontroller}
			u(t) = K x(t) + L r(t)
		\end{align}
		ensures that $\lim_{t \to \infty} e(t) = 0$, where $L = \Gamma - K \Pi$.		

	\section{MPC controller for tracking}
	In this section, MPC method for tracking bounded references with arbitrary dynamics is presented. 
	The dynamics of state error is analyzed first. 
	Then, required properties of cost function and terminal constraint are presented.
	Finally, the proposed MPC method is concluded and its properties are proved theoretically.
	
	\subsection{Error Dynamics}
	Consider that 
	\begin{align} \label{eqn:eCxtilde}
		\begin{aligned}
			e(t) &=   Cx(t) - Q^e r(t) \\
			&=   C \big( x(t)-\Pi r(t) \big) + (C \Pi - Q^e) r(t) \\
			&=   C \big( x(t)-\Pi r(t) \big).
		\end{aligned}
	\end{align}
	
	Notate $\tilde{x}(t) = x(t)-\Pi r(t)$, which is referred to as state error. Then, $e(t) = C \tilde{x}(t)$.

	Consider the control law which is given as
	\begin{align}	\label{eqn:uKxLrv}
		u(t) = K x(t) + L r(t) + v(t),
	\end{align}
	where the free control variable $v(t)$ is  the difference between actual control input $u(t)$ and the feedback controller (\ref{eqn:IMPcontroller}).
	It can be rewritten as 
	\begin{align}	\label{eqn:uKxtildeGammarv}
		\begin{aligned}
			u(t) &= K \big( x(t)-\Pi r(t) \big) + (K \Pi + L) r(t) + v(t) \\
			&= K \tilde{x}(t) + \Gamma r(t) + v(t).
		\end{aligned}
	\end{align}
	
	With control law (\ref{eqn:uKxLrv}), it is derived that
	\begin{align}	\label{eqn:xtilderABKxtildeBv}
			\tilde{x}(t+1) &= x(t+1) - \Pi r(t+1) 	\nonumber \\
			&= Ax(t) + B \big( K \tilde{x}(t) + \Gamma r(t) + v(t) \big) - \Pi S r(t)  \nonumber	\\
			&= A_\mathrm{cl} \tilde{x}(t)   + (A \Pi + B \Gamma -\Pi S) r(t)   	+ Bv(t)	\nonumber \\
			&= A_\mathrm{cl} \tilde{x}(t) + Bv(t).
	\end{align}

	Then, the tracking problem can be solved by solving the stabilization problem of state error $\tilde{x}(t)$ with control input $v(t)$. 
	In the following sections, MPC controller is designed to ensure the asymptotic stability of $\tilde{x}(t)$ while maintaining constraint satisfaction.

\subsection{Cost Function}
	To bring in flexibility in dealing with the given reference $r(t)$, artificial reference $\bar{r}(t)$ is introduced.
	The deviation between the actual reference $r(t)$ and artificial reference $\bar{r}(t)$ is  $e_r(t) = {r}(t) - \bar{r}(t)$,  
		which is referred to as reference error. 
	The artificial state error is defined as 
	\begin{align}
		\bar{x}(t) = x(t) - \Pi \bar{r}(t).
	\end{align}
	
	It is found that 
	\begin{align} \label{eqn:xtilde_xbar_er}
		\tilde{x} = \bar{x}(t) - \Pi e_r(t).
	\end{align}
	
	Thus, convergence of state error $\tilde{x}$ is achieved when both the artificial state error ${\bar{x}}(t)$ and reference error $e_r(t)$ converge to 0.
	Thus, ${\bar{x}}(t)$ and $e_r(t)$ should be penalized.
	
	Notate the predicted control variable and predicted artificial reference at $t+k$ as 
	${{v}}(k|t)$ and $\bar{r}(k|t)$, $k \in \mathbb{N}_{[0,N]}$, 	where $N$ is the prediction horizon. 
	Notate 
	\begin{align}
		\bm{{v}}(t)     &= \begin{bmatrix} {{v}}'(0|t)     & {{v}}'(1|t)     & \dots & {{v}}'(N-1|t) \end{bmatrix}', \\
		\bm{\bar{r}}(t) &= \begin{bmatrix} {\bar{r}}'(0|t) & {\bar{r}}'(1|t) & \dots & {\bar{r}}'(N|t) \end{bmatrix}',
	\end{align}	
	where ${\bar{r}}(0|t) = \bar{r}(t)$ and ${\bar{r}}(k+1|t) = S {\bar{r}}(k|t)$.
	With $\bm{{v}}(t)$ and $\bm{\bar{r}}(t)$, the predicted artificial state error ${\bar{x}}(k|t)$ is determined correspondingly, 
		where $\bar{x}(0|t) = x(t) - \Pi \bar{r}(0|t), 
			   \bar{x}(k+1|t) = A_\mathrm{cl} \bar{x}(k|t) + B v(k|t)$.
	Notate 
	\begin{align}
		\bm{\bar{x}}(t) &= \begin{bmatrix} {\bar{x}}'(0|t) & {\bar{x}}'(1|t) & \dots & {\bar{x}}'(N|t) \end{bmatrix}'.
	\end{align}
	
	Then, cost function is given as
	\begin{align}
		J( \bm{\bar{x}}(t), e_r(t)  ) = J_{\bar{x}} ( \bm{\bar{x}}(t) ) + J_{\bar{r}} (e_r(t) ),
	\end{align}
	where
	\begin{align}
		J_{\bar{x}} ( \bm{\bar{x}}(t) )  &=  	
		\sum_{k=0}^{N-1} \left \| \bar{x}(k|t) \right \| _Q^2 
		+  \left \| \bar{x}(N|t) \right \| _P^2, \\
		J_{\bar{r}} (e_r(t) )  &= \left \| e_r(t) \right \| _T^2.
	\end{align}
	
	$Q \in \mathbb{R}^{n \times n}$, $P \in \mathbb{R}^{n \times n}$, 
	and $T \in \mathbb{R}^{q \times q}$ are weight matrices which satisfy the following assumptions.
	
	\textbf{(A6)} $Q$ and $P$ are symmetrical and positive definite matrices which satisfy $A_\mathrm{cl}'P A_\mathrm{cl} -  P + Q = 0$.

	\textbf{(A7)} $T$ is a symmetrical and positive definite matrix which satisfies $S'TS = T$.

	(A6) requires that $A_\mathrm{cl}, Q$, and $P$ satisfy the discrete-time Lyapunov equation.
	Since $A_\mathrm{cl}$ is Schur, given a symmetrical and positive definite matrix $Q$, 
	a unique $P$ can be determined which is symmetrical and positive definite, as well.
	Weight matrix $T$ satisfying (A7) will be determined in Section IV.
		
\subsection{Terminal Constraint}
	Similar as (\ref{eqn:xtilderABKxtildeBv}), with control input given as $u(t) = K x(t) + L \bar{r}(t) + v(t)$, 
	it can be derived that 
	\begin{align}
		&u(t)  =  K \bar{x}(t) + \Gamma \bar{r}(t) + v(t), \\
		&\bar{x}(t+1) = A_\mathrm{cl} \bar{x}(t) + Bv(t).
	\end{align}
	
	With $v(t)=0$, the augmented system of artificial state error and artificial reference and constraints (\ref{eqn:systemConstraints})  are rewritten as
		\begin{align}
				&\begin{bmatrix} \bar{x}(t+1) \\ \bar{r}(t+1)  \end{bmatrix} = 
				\begin{bmatrix} A_\mathrm{cl} & 0  \\  0 & S  \end{bmatrix}
				\begin{bmatrix} \bar{x}(t) \\ \bar{r}(t)   \end{bmatrix}, 		\label{eqn:augmented_system}\\ 
				&\begin{bmatrix} I & \Pi  \\  K & \Gamma \end{bmatrix}
				\begin{bmatrix} \bar{x}(t)  \\  \bar{r}(t) \end{bmatrix} \in \mathcal{Z}. \label{eqn:constraint_augmented_system}
			\end{align}
	
	The terminal constraint, notated as $\begin{bmatrix} \bar{x}'  &  \bar{r}' \end{bmatrix}' \in \mathcal{Z}_f $, is supposed to satisfy the following assumption.
	
	\textbf{(A8)}
	For all $\begin{bmatrix} \bar{x}'  &  \bar{r}' \end{bmatrix}' \in \mathcal{Z}_f$, 
	
	(i)  $\begin{bmatrix} A_\mathrm{cl} & 0  \\  0 & S  \end{bmatrix} \begin{bmatrix} \bar{x}  \\  \bar{r} \end{bmatrix} \in \mathcal{Z}_f $,
	(ii) $\begin{bmatrix} I & \Pi  \\  K & \Gamma \end{bmatrix}
	       \begin{bmatrix} \bar{x}  \\  \bar{r} \end{bmatrix} \in \mathcal{Z}$.
	
	Assumption (A8) is presented separately here for ease of reference. 
	Terminal constraint $\mathcal{Z}_f$ which satisfies (A8) will be determined in Section IV.
	
	For all $\begin{bmatrix} \bar{x}'(0)  &  \bar{r}'(0) \end{bmatrix}' \in \mathcal{Z}_f $, 
		with $u(t)  =  K \bar{x}(t) + \Gamma \bar{r}(t)$, 
		we have $\begin{bmatrix}x'(t) & u’(t)\end{bmatrix}' \in \mathcal{Z}, t \in \mathbb{N}_0^+$.
	Meanwhile, since $A_\mathrm{cl}$ is Schur, we have $\lim_{t \to \infty} \bar{x}(t) = 0$ and correspondingly, 
		$\lim_{t \to \infty} C \bar{x}(t) - Q^e \bar{r}(t)= 0$.
	This means, system (\ref{eqn:actualSystem}) under constraint (\ref{eqn:systemConstraints}) can track reference $\bar{r}(t)$ without violating constraints.
	Such kind of references is said to be admissible.
	The set of admissible references is notated as
	\begin{align}		\label{eqn:Rf_gen}
		\mathcal{R}_{f} = \left \{
		\bar{r} | \exists \bar{x} \text{ such that } 	\begin{bmatrix} \bar{x}'  &  \bar{r}' \end{bmatrix}' \in \mathcal{Z}_f
		\right \}.
	\end{align}
	
	Assumption (A8) implies that if  $\begin{bmatrix} \bar{x}'(t)   &  \bar{r}'(t) \end{bmatrix}' \in \mathcal{Z}_f $,
	then $\begin{bmatrix} (A_\mathrm{cl}\bar{x})' &  (S\bar{r})' \end{bmatrix}' \in \mathcal{Z}_f $.
	Thus, $ \bar{r} \in \mathcal{R}_{f}$ implies  $S \bar{r} \in \mathcal{R}_{f}$ and further, $S \mathcal{R}_{f} \subseteq \mathcal{R}_{f}$.
	
\subsection{Proposed MPC Controller}
	The proposed MPC controller is given as
	\begin{align}		\label{eqn:MPCcontroller}
		\pi(x(t),r(t)) &= K \bar{x}^*(0|t) + \Gamma \bar{r}^*(0|t) + v^*(0|t),  		
	\end{align}
	where $\bar{x}^*(0|t)$, $\bar{r}^*(0|t)$ and $v^*(0|t)$ are determined by the quadratic optimization problem $\mathbb{QP}$ given as
	\begin{align}		\label{eqn:QP}
		\min_{\bar{r}(t), \bm{v}(t)} ~ 	J \big( \bm{\bar{x}}(t), e_r(t)   \big)
	\end{align}
	subject to 	
	\begin{subequations}  \label{eqn:QP_constraints}
		\begin{align}
			\bar{x}(0|t)   = x(t) - \Pi \bar{r}(0|t), \bar{r}(0|t) = \bar{r}(t), \label{eqn:QP_initialCondition}\\
			\begin{bmatrix} \bar{x}(k+1|t) \\ \bar{r}(k+1|t)  \end{bmatrix} = 
			\begin{bmatrix} A_\mathrm{cl} & 0  \\  0 & S  \end{bmatrix}
			\begin{bmatrix} \bar{x}(k|t) \\ \bar{r}(k|t)   \end{bmatrix}
			+\begin{bmatrix} Bv(k|t) \\ 0   \end{bmatrix}, \\
			\begin{bmatrix} I & \Pi  \\  K & \Gamma \end{bmatrix}
			\begin{bmatrix} \bar{x}(k|t)  \\  \bar{r}(k|t) \end{bmatrix} + 
			\begin{bmatrix} 0 \\ v(k|t)  \end{bmatrix}		\in \mathcal{Z} ,		\label{eqn:QP_feasibleConstraints}\\
			\begin{bmatrix} \bar{x}'(N|t)  &  \bar{r}'(N|t) \end{bmatrix}' \in \mathcal{Z}_f, 		\label{eqn:QP_terminalConstraints}\\
			k \in \mathbb{N}_0^{N-1}.   \label{eqn:QP_counts}
		\end{align}
	\end{subequations}

	The constraint satisfaction, recursive feasibility, and asymptotic stability properties of the proposed MPC controller (\ref{eqn:MPCcontroller}) are concluded in Theorem 1.

	\textbf{Theorem 1:}
	Consider system (\ref{eqn:actualSystem}) and reference (\ref{eqn:referenceSystem}) satisfying (A1) to (A5).
	The controller is given as (\ref{eqn:MPCcontroller}) with weight matrices $Q,P,T$ satisfying (A6) and (A7) and terminal constraint $\mathcal{Z}_f$ satisfying (A8).
	If $\mathbb{QP}$  is feasible at $t$, then
	
	(i)    (constraint satisfaction) constraint (\ref{eqn:systemConstraints}) is satisfied;
	
	(ii)   (recursive feasibility) $\mathbb{QP}$  is feasible at $t+1$;
	
	(iii)  (asymptotic stability) if $r(t) \in \mathcal{R}_f^\mathrm{m}$, $\bar{x}^*(0|t)$ converges to 0, $\bar{r}^*(0|t)$ converges to $r(t)$, $e(t)$ converges to 0, where 
	\begin{align*}
		\mathcal{R}_f^\mathrm{m} = \left\{ r \Big|
			\begin{bmatrix} I & \Pi  \\  K & \Gamma \end{bmatrix}
			\begin{bmatrix} 0  \\  S^k r \end{bmatrix} \in \mathcal{Z},
			k \in \mathbb{N}_0^{N-1},		
			\begin{bmatrix} 0  \\  S^N r \end{bmatrix} \in \mathcal{Z}_f \right\}.
	\end{align*}
	
	\begin{proof}
		(i) This is ensured by (\ref{eqn:QP_feasibleConstraints}) with $k=0$.
		
		(ii) 
		Suppose the optimal solution is achieved with $\bar{r}^*(t)$ and $\bm{v}^*(t) = 
		\begin{bmatrix}
			v^{*'}(0|t), v^{*'}(1|t), \cdots, v^{*'}(N-1|t)
		\end{bmatrix}'$. 
		Then, the predicted artificial references and the corresponding predicted artificial state error can be determined correspondingly and are notated as
		\begin{align}
			\bm{\bar{r}}^*(t) &= 
			\begin{bmatrix}
				\bar{r}^{*'}(0|t), \bar{r}^{*'}(1|t), \cdots, \bar{r}^{*'}(N|t)
			\end{bmatrix}', \\
			\bm{\bar{x}}^*(t) &= 
			\begin{bmatrix}
				\bar{x}^{*'}(0|t), \bar{x}^{*'}(1|t), \cdots, \bar{x}^{*'}(N|t)
			\end{bmatrix}'.
		\end{align}
		
		At $t+1$,  $x(t+1) = \bar{x}^{*}(1|t) + \Pi \bar{r}^{*}(1|t)$ and $r(t+1) = Sr(t)$.
		With $\bar{r}^\dagger(t+1) = S\bar{r}^*(t)$  and
		$\bm{v}^\dagger(t+1)= 	
		\begin{bmatrix} v^{*'}(1|t), \cdots, v^{*'}(N-1|t), 0 \end{bmatrix}'$,
		the predicted artificial references and the corresponding predicted artificial state error at $t+1$ can be determined correspondingly and are given as
		\begin{align}
			&\bm{\bar{r}}^\dagger(t+1) = 
			\begin{bmatrix}
				\bar{r}^{*'}(1|t), \cdots, \bar{r}^{*'}(N|t), \big(S \bar{r}^*(N|t) \big)'
			\end{bmatrix}', \\
			&\bm{\bar{x}}^\dagger(t+1) = 
			\begin{bmatrix}
				\bar{x}^{*'}(1|t), \cdots, \bar{x}^{*'}(N|t), \big( A_\mathrm{cl}\bar{x}^*(N|t) \big)'
			\end{bmatrix}'.
		\end{align}
		
		According to  (A8), (\ref{eqn:QP_terminalConstraints}) implies 
					$\begin{bmatrix} \bar{x}'(N|t)  &  \bar{r}'(N|t) \end{bmatrix}' $ $\in \mathcal{Z} $ and
			$\begin{bmatrix} \big( A_\mathrm{cl}\bar{x}(N|t) \big)' &  \big( S\bar{r}(N|t) \big)' \end{bmatrix}' \in \mathcal{Z}_f $.
		Along with (\ref{eqn:QP_feasibleConstraints}), it can be found that the solution with $\bar{r}^\dagger(t+1)$ and $\bm{\bar{v}}^\dagger(t+1)$ is feasible to $\mathbb{QP}$ at $t+1$.
		Thus, (ii) holds.		
		
		(iii) 
		Notate 
		\begin{align}
			J^*(t) = J \big( \bm{\bar{x}}^*(t), e_r^*(t) \big), 
			J^\dagger(t) = J \big( \bm{\bar{x}}^\dagger(t), e_r^\dagger(t) \big).
		\end{align}
		
		With the feasible solution given in (ii), we have 
		
		\begin{align}
			\begin{aligned}
				&	  J^*(t+1) - J^*(t) \leq J^\dagger(t+1) - J^*(t)\\
				=&     \left \| A_\mathrm{cl}\bar{x}^*(N|t) \right \| _P^2 + \left \| \bar{x}^*(N|t) \right \| _Q^2  - \left \| \bar{x}^*(N|t) \right \| _P^2  \\
				+& \left \| S\bar{r}^*(t) - r(t+1) \right \| _T^2 - \left \| \bar{r}^*(t) - r(t) \right \| _T^2   \\
				-& \left \| \bar{x}^*(0|t) \right \| _Q^2.
			\end{aligned}
		\end{align}
		
		According to (A6) and (A7), we have 
		\begin{align}
			&\begin{aligned}
				& \left \| A_\mathrm{cl}\bar{x}^*(N|t) \right \| _P^2 + \left \| \bar{x}^*(N|t) \right \| _Q^2 - \left \| \bar{x}^*(N|t) \right \| _P^2\\
				=   &\left \| \bar{x}^*(N|t) \right \| _{A_\mathrm{cl}'PA_\mathrm{cl} -  P + Q}^2 = 0.
			\end{aligned} \\
			&\begin{aligned}
				&\left \| S\bar{r}^*(t) - r(t+1) \right \| _T^2 - \left \| \bar{r}^*(t) - r(t) \right \| _T^2 \\
				=   &\left \| S\bar{r}^*(0|t) - Sr(t) \right \| _T^2 - \left \| \bar{r}^*(0|t) - r(t) \right \| _T^2  \\
				=   &\left \|  \bar{r}^*(0|t) -  r(t) \right \| _{S'TS-T}^2 = 0.
			\end{aligned}
		\end{align}
		
		Then, $J^*(t+1) - J^*(t) \leq  - \left \| \bar{x}^*(0|t) \right \| _Q^2$.
		Since $Q,P$, and $T$ are all positive definite, 
		$ J( t)$ is non-negative and non-increasing.
		Thus, $J(t)$ converges to a constant and $ \bar{x}^*(0|t)$ converges to $0$.
		When $ \bar{x}^*(0|t) = 0$, it is found that if $r(t) \in \mathcal{R}_f^\mathrm{m}$, 
		the solution with $\bar{r}(t) = r(t)$ and $\bm{{v}}(t) = 0$ is feasible and the corresponding cost is $J(t) = 0$, which is the globally optimal.
		Thus, $\bar{r}^*(0|t)$ converges to $r(t)$, $v^*(0|t)$ converges to $0$, and $u(t)$ converges to $Kx(t)+Lr(t)$. 
		According to Lemma 1, $e(t)$ converges to $0$.
		Thus, (iii) holds.
	\end{proof}
	
		
		\textbf{Remark 2:}
		In Theorem 1, assumptions (A1)-(A8) are needed.
		Among those assumptions, (A1), (A2),(A3), and (A5) are necessary conditions for constrained tracking. 
		(A4)  indicates that the reference to be tracked is bounded.
		(A6) can be easily satisfied by solving the discrete-time Lyapunov equation to get $P$ with the given $A_\mathrm{cl}, Q$.
		For physical applications, weight matrix $T$ and terminal constraint $\mathcal{Z}_f$ satisfying (A7) and (A8) are determined in Section IV.		
	
		
		\textbf{Remark 3:}
		It can be seen that $r(t)$ is not included in constraints (\ref{eqn:QP_constraints}), thus, 
			the region of attraction domain, that is, the set of $x(t)$ such that (\ref{eqn:QP}) has a feasible solution, is independent of reference $r(t)$.
		Then, (ii) holds even when reference switches sometimes.
		Meanwhile, as long as (\ref{eqn:referenceSystem}) holds when $t \in \mathbb{N}_{t_1}^{t_2}$, $e(t_2)$ becomes sufficiently small if $t_2 -t_1$ is sufficiently large.


\section{Periodic and Non-periodic cases}

The proposed MPC controller ensures the convergence of tracking error as long as weight matrices and terminal constraint satisfy (A6) to (A8).
In this section, periodic and non-periodic references are considered respectively 
to show the existence of weight matrix and terminal constraint satisfying required assumptions.

\subsection{Periodic case}
	Consider that the given reference is periodic. 
	Then, there exists $k_0 \in \mathbb{N}^+$ such that 
	\begin{align} \label{eqn:Sk0I}
		S^{k_0} = I.
	\end{align}
	
	In this case, it is easy to find a symmetrical and positive definite matrix $T$ satisfying (A7), which is given in the following lemma.
	
		
		\textbf{Lemma 2:}
		If (\ref{eqn:Sk0I}) holds, $T = \sum_{i=1}^{k_0} {S^i}' T_0 S^i$ with a symmetrical and positive definite matrix $T_0$ satisfies (A7).

	\begin{proof}
		Since $T_0$ is symmetrical and positive definite, ${S^i}' T_0 S^i$ is symmetrical and positive definite for all positive integer $i$.
		Further, $T = \sum_{i=1}^{k_0} {S^i}' T_0 S^i$ is symmetrical and positive definite, as well. 
		Meanwhile, 
		\begin{align}
			\begin{aligned}
			S'T S  
				   &= \sum_{i=1}^{k_0-1} (S^i S)' T_0 (S^i S) + (S^{k_0} S)' T_0 S^{k_0} S \\
				   &= \sum_{i=2}^{k_0} {S^i}' T_0 S^i + {S}' T_0 S  
				    = \sum_{i=1}^{k_0} {S^i}' T_0 S^i  = T.
			\end{aligned}
		\end{align}
		
	Thus, Lemma 2 holds.
	\end{proof}
	
	$\mathcal{Z}_f$ can be chosen as the maximal output admissible set $\mathcal{O}_\infty$ introduced in \cite{gilbert1991linear}.
	Define $t$-step output admissible set as
			\begin{align}
		&\mathcal{O}_0 =  \left\{
		\begin{bmatrix}	\bar{x}'  &  \bar{r}'	\end{bmatrix}'  |
		\begin{bmatrix}	\bar{x}'  &  \bar{r}'	\end{bmatrix}' \in {\mathcal{Z}}		
		\right\} , \\
		&\begin{aligned}    \label{eqn:Ot_t1}
			\mathcal{O}_t =  \left\{
			\begin{bmatrix}	\bar{x}  \\  \bar{r}	\end{bmatrix} \Big |			
			\begin{bmatrix} I & \Pi  \\  K & \Gamma \end{bmatrix}
			\begin{bmatrix} \bar{x}  \\  \bar{r} \end{bmatrix} \in {\mathcal{Z}},
			\begin{bmatrix} A_\mathrm{cl} \bar{x} \\ S \bar{r}  \end{bmatrix} \in \mathcal{O}_{t-1}
			\right\}.
		\end{aligned}
	\end{align}

	$\mathcal{O}_\infty$ is determined when $t$ goes to infinity.
	If there exists an integer $t_0$ such that $\mathcal{O}_{t_0+1} = \mathcal{O}_{t_0}$, 
	then, $\mathcal{O}_{t+1} = \mathcal{O}_{t} =  \mathcal{O}_{t_0}, \forall t\geq t_0$ and further, $\mathcal{O}_{\infty} =  \mathcal{O}_{t_0}$.
	This means, $\mathcal{O}_{\infty}$ is determined in $t_0$ steps.
	As shown in \cite{gilbert1991linear}, when $A_\mathrm{cl}$ and $S$ are both Schur, 
		$\mathcal{O}_{\infty}$ can be determined in finite steps
 	Fortunately, when reference is periodic, that is, $S^{k_0} = I$, it can be proved that $\mathcal{O}_\infty$ can be determined in finite steps as well,
	which is given in the following lemma.
	
		
		\textbf{Lemma 3:}
		If (\ref{eqn:Sk0I}), (A2), and (A3) holds, then 
		(i) $\mathcal{O}_\infty$ corresponding to system (\ref{eqn:augmented_system}) under constraint (\ref{eqn:constraint_augmented_system}) can be determined in finite steps,
		(ii) $\mathcal{Z}_f = \mathcal{O}_\infty$ satisfies (A8).

	\begin{proof}
		(i) Consider the constrained system given as
		\begin{align} \label{eqn:temp_dafeqdaf}
			\begin{bmatrix}	\bar{x}(t+1) \\ \bm{w}(t+1) \end{bmatrix} = 
			\begin{bmatrix}	A_\mathrm{cl} & 0 \\ 0 & I	\end{bmatrix} 
			\begin{bmatrix}	\bar{x}(t) \\ \bm{w}(t) \end{bmatrix}, 
			\begin{bmatrix}	\bar{x}(t) \\ \bm{w}(t) \end{bmatrix} \in \Omega,
		\end{align}
		where $\bm{w}(t) = \begin{bmatrix}	w_1'(t) & w_2'(t)& ... & w_{k_0-1}(t)	\end{bmatrix}'$, $w_i(t)\in \mathbb{R}^{q}$, $i\in \mathbb{N}_1^{\bar{q}}$,
		and $ \Omega$ is given as
		\begin{align}
			 \Omega = \left\{
			 \begin{bmatrix} \bar{x}(t) \\ \bm{w}(t) \end{bmatrix} \Big|
			 \begin{bmatrix} I & \Pi  \\  K & \Gamma \end{bmatrix}
			 \begin{bmatrix} A_\mathrm{cl}^{i-1} \bar{x}(t) \\ w_i(t) \end{bmatrix} \in \mathcal{Z},
			 i\in \mathbb{N}_1^{\bar{q}}
			 \right\}.
		\end{align}
		
		As shown in \cite{gilbert1991linear} and \cite{gilbert2011constrained}, 
		the maximal output admissible set $\mathcal{O}^{xw}_\infty$ for system (\ref{eqn:temp_dafeqdaf}) can be finitely determined. 
		Then, 
		\begin{align} 	\label{eqn:tempjlakoui}
		\begin{aligned}
			\mathcal{O}_\infty = 
			&\left\{
				\begin{bmatrix} \bar{x}'  &  \bar{r}' \end{bmatrix}’ \Big|
				\begin{bmatrix}	\bar{x}' &  \bm{w}' \end{bmatrix}' \in \mathcal{O}^{zw}_\infty, 
			\right. \\
			&\left.				
				\bm{w} =  \begin{bmatrix}	\bar{r}' & (S \bar{r})'  & ( S^2 \bar{r} )' & ... & (S^{k_0-1} \bar{r})'	\end{bmatrix}'
			\right\}.
		\end{aligned}
		\end{align}
		This means, $\mathcal{O}_\infty$ can be determined according to (\ref{eqn:tempjlakoui}) with finitely determined $\mathcal{O}^{xw}_\infty$ and (i) holds.	
			
	(ii) holds according to definition  of  $O_\infty^{xw}$ and $O_\infty$.
	\end{proof}

	With the analysis given above, with $T = \sum_{i=1}^{k_0} {S^i}' T_0 S^i$ and $\mathcal{Z}_f = \mathcal{O}_\infty$, MPC controller (\ref{eqn:MPCcontroller}) can be conducted.
	Additional properties of $\mathcal{R}_f$ and $\mathcal{R}_f^\mathrm{m}$ are given below.
	
		
		\textbf{Lemma 4:}
		With  $\mathcal{Z}_f = \mathcal{O}_\infty$, the following properties hold:
		(i)     $\mathcal{R}_f = S\mathcal{R}_f$;
		(ii)    for all $r\in \mathcal{R}_f$, $\begin{bmatrix} 0  &  r' \end{bmatrix}' \in \mathcal{Z}_f$;
		(iii)	$\mathcal{R}_{f} = \left \{	r |
					 \begin{bmatrix} 0  &  r' \end{bmatrix}' \in \mathcal{Z}_f	\right\}$,
		(iv)   $\mathcal{R}_f^\mathrm{m} = \mathcal{R}_f$.
	
	\begin{proof}
		When $k_0 = 1$, (i) is obviously true. 
		When $k_0 \in \mathbb{N}^+$ is greater than $1$, 
		$S \mathcal{R}_f \subseteq \mathcal{R}_f$ and further,  $S^{k_0} \mathcal{R}_f \subseteq S \mathcal{R}_f$.
		Meanwhile, $S^{k_0} = I$.
		Then, $\mathcal{R}_f = S^{k_0} \mathcal{R}_f \subseteq S \mathcal{R}_f$.
		Thus, $\mathcal{R}_f = S \mathcal{R}_f$ and (i) holds.
		
		For all $r\in \mathcal{R}_f$, there exists $\bar{x}$ such that $\begin{bmatrix} \bar{x}'  &  r' \end{bmatrix}' \in \mathcal{Z}_f$.
		According to the property of $\mathcal{Z}_f$, 
		for all $i \in \mathbb{N}$, $\begin{bmatrix} (A_\mathrm{cl}^i \bar{x})'  &  (S^ir)' \end{bmatrix}' \in \mathcal{Z}_f$.
		Then, 		
		\begin{align}
			    \lim_{j \to \infty} \begin{bmatrix} (A_\mathrm{cl}^{k_0 j} \bar{x})'  &  (S^{k_0 j} r)' \end{bmatrix}'  
			=   \lim_{j \to \infty} \begin{bmatrix} 0  &  (S^{k_0 j} r)' \end{bmatrix}' 
			\in \mathcal{Z}_f  \nonumber
		\end{align}
		
		Since  $S^{k_0} = I$, $\begin{bmatrix} 0  &  r' \end{bmatrix}' = \lim_{j \to \infty} \begin{bmatrix} 0  &  (S^{k_0 j} r)' \end{bmatrix}'  \in \mathcal{Z}_f$.
		Thus, (ii) holds.
		
		According to (ii), (iii) holds.
		
		 According to (ii), for all $r \in \mathcal{R}_f$,  $\begin{bmatrix} 0  &  r' \end{bmatrix}' \in \mathcal{Z}_f$.
		 According to the property of $\mathcal{Z}_f$, 
		 $\begin{bmatrix} I & \Pi  \\  K & \Gamma \end{bmatrix}
		 \begin{bmatrix} 0  \\  S^k r \end{bmatrix} \in \mathcal{Z},	
		 \begin{bmatrix} 0  \\  S^k r \end{bmatrix} \in \mathcal{Z}_f,
		 k \in \mathbb{N}_0^N$.
		 Thus, $\mathcal{R}_f \subseteq \mathcal{R}_f^\mathrm{m}$.
		 
		 For all $r \in \mathcal{R}_f^\mathrm{m}$, 		 
		 $\begin{bmatrix} 0  &  (S^Nr)' \end{bmatrix}' \in \mathcal{Z}_f$.
		 Thus, $S^Nr \in \mathcal{R}_f$. 
		 According to property (i), $r \in S^{-N}\mathcal{R}_f = S^{-N} S^N \mathcal{R}_f= \mathcal{R}_f$.
		 Thus, $\mathcal{R}_f^\mathrm{m} \subseteq \mathcal{R}_f$.
		 
		 Thus, $\mathcal{R}_f^\mathrm{m} = \mathcal{R}_f$ and (iv) holds.
	\end{proof}

\subsection{non-periodic case}
	Consider that the given reference is non-periodic. Then, 
	\begin{align} 
		S^{k} \neq I, \forall k\in \mathbb{N}^{+}.
	\end{align}

	Non-periodic reference $r$ contains periodic and non-periodic elements.  Without loss of generality,
	$r$ is expressed as $r(t) = \begin{bmatrix}	\alpha'(t) & \beta'(t)	\end{bmatrix}'$,
	where $\alpha(t) \in \mathbb{R}^{q_\mathrm{p}}$ is periodic, $\beta(t) \in \mathbb{R}^{q_\mathrm{n}}$ is non-periodic, and $q_\mathrm{p} + q_\mathrm{n} = q$.
	Further, $S = diag(S_\mathrm{p}, S_\mathrm{n})$ where $S_\mathrm{p} \in \mathbb{R}^{q_\mathrm{p} \times q_\mathrm{p}}, S_\mathrm{n} \in \mathbb{R}^{q_\mathrm{n} \times q_\mathrm{n}}$.
	There exists $k_0$ such that $S_\mathrm{p} ^{k_0} = I$ and $S_\mathrm{n}^{k} \neq I, \forall k\in \mathbb{N}^{+}.$

	Assumption (A4) implies that eigenvalues of $S$ include $\pm 1$ and conjugate complex numbers whose amplitudes are $1$.
	Moreover, the Lyapunov stability of exosystem (\ref{eqn:referenceSystem}) implies that the algebraic multiplicity of $\pm 1$ is the same as the geometric multiplicity. 
	Thus, $S$, as well as $S_\mathrm{p}$ and $S_\mathrm{n}$, is diagonalizable. 
	Without loss of generality, $S_\mathrm{n}$ is supposed to have a block diagonal form, which is expressed as
	$S_\mathrm{n} = diag(S_1, S_2, ..., S_{\bar{q}})$, where $S_i \in \mathbb{R}^{2 \times 2}, 
	\lambda(S_i)\in \mathbb{C},\left | \lambda(S_i)  \right | = 1, i \in \mathbb{N}_1^{\bar{q}}$.	
	
	Since $r(t)$ is non-periodic,  weight matrix $T$ cannot be determined according to Lemma 2.
	In this case, $T$ satisfying (A7) is given in the following lemma.
	
		
		\textbf{Lemma 5:}
		The block diagonal matrix $T=diag(T_\mathrm{p},T_\mathrm{n})$ satisfies (A7), where
		$T_\mathrm{p} =  \sum_{i=1}^{k_0} {S_\mathrm{p}^i}' T_0 S_\mathrm{p}^i$ with a symmetrical and positive definite matrix $T_0$ 
			and $T_\mathrm{n} = diag(T_1, T_2, ..., T_{\bar{q}})$.
		$T_i=\Lambda_{T_i} (E_i E_i^H)^{-1}$, where 
		$\Lambda_{T_i} \in \mathbb{R}^{2 \times 2}$ is positive definite and diagonal, 
		$E_i \in \mathbb{C}^{2 \times 2}$ is a complex matrix such that $S_i = E_i \Lambda_i E^{-1}_i,
		\Lambda_i = diag(\lambda_i, \mathring{\lambda_i})$, and  $\lambda_i$ is an eigenvalue of $S_i$.

	\begin{proof}
		According to Lemma 2, $T_\mathrm{p}$ is  symmetrical and positive definite matrix which satisfies $S_\mathrm{p}' T_\mathrm{p} S_\mathrm{p} = T_\mathrm{p}$.	
		
		According to (A4), 
		$S_i$ is diagonalizable and there exists $E_i \in \mathbb{C}^{2 \times 2}$ 
		such that $S_i = E_i \Lambda_i E^{-1}_i,
		\Lambda_i = diag(\lambda_i, \mathring{\lambda_i})$, 
		where $\lambda_i$ and $\mathring{\lambda_i}$ are eigenvalues of $S_i$.
		Meanwhile, according to (A4), we have$|\lambda_i| = |\mathring{\lambda_i}| = 1$.
		Then,
		\begin{align}
			  \Lambda_i^H {\Lambda}_i  
			= \begin{bmatrix}	\lambda_i\mathring{\lambda_i} & 0 \\ 0 & \lambda_i\mathring{\lambda_i}	\end{bmatrix}
			= \begin{bmatrix}	|\lambda_i|^2 & 0 \\ 0 & |\lambda_i|^2	\end{bmatrix} = I.
		\end{align}
	
		$E_i$ can be given as $E_i = \begin{bmatrix}	u & \mathring{u} \\ v &\mathring{v}	\end{bmatrix}$.
		Then,
		\begin{align}
			E_i E_i^H = \begin{bmatrix}	u & \mathring{u} \\ v &\mathring{v}	\end{bmatrix}
			\begin{bmatrix}	\mathring{u} & \mathring{v}	\\  u & v\end{bmatrix} 
			= \begin{bmatrix}	2u\mathring{u} & u\mathring{v}+\mathring{u}v \\  
				u\mathring{v}+\mathring{u}v & 2v\mathring{v}	\end{bmatrix}.
		\end{align}
		$u\mathring{u} = |u|^2$ and $v\mathring{v} = |v|^2$ are real numbers.
		Since $u\mathring{v}$ and $\mathring{u}v$ are conjugate, $u\mathring{v}+\mathring{u}v$ is a real number.
		Thus, $E_i E_i^H \in \mathbb{R}^{2 \times 2}$.
		Further, $(E_i E_i^H)^{-1} \in \mathbb{R}^{2 \times 2}$ and $T_i=\Lambda_{T_i} (E_i E_i^H)^{-1} \in \mathbb{R}^{2 \times 2}$.
		
		Since $E_i E_i^H \in \mathbb{R}^{2 \times 2}$, we have $(E_i E_i^H)' = (E_i E_i^H)^H = E_i E_i^H$. Thus, $E_i E_i^H$ as well as $(E_i E_i^H)^{-1}$ is symmetric.

		Meanwhile, it is found that 
		\begin{align} 
		\begin{aligned}
			\mathrm{det}(E_i E_i^H) &= 4 u \mathring{u} v\mathring{v}
			- ( u\mathring{v}+\mathring{u}v )  ( u\mathring{v}+\mathring{u}v )       \\
			&= 2 u \mathring{u} v\mathring{v} - (u\mathring{v})^2 - (\mathring{u}v)^2 \\
			&= -(u\mathring{v} - \mathring{u}v)^2.
		\end{aligned}
		\end{align}
		Since $u\mathring{v}$ and $\mathring{u}v$ are conjugate, $u\mathring{v} - \mathring{u}v$ is on the imaginary axis.
		Thus, $\mathrm{det}(E_i E_i^H) = -(u\mathring{v} - \mathring{u}v)^2$ is positive.
		Meanwhile, $2u\mathring{u} = 2|u|>0$.
		Thus, $E_i E_i^H$ is positive definite.
		Further, $(E_i E_i^H)^{-1}$ is positive definite, as well.
		
		Since $\Lambda_{T_i}$ is a positive and diagonal matrix, 
		$\Lambda_{T_i} E_i E_i^H$ is symmetrical and positive definite.
		Meanwhile,
		\begin{align}
		\begin{aligned}
			S_i' T_i S_i &= S_i^H T_i S_i \\
					   	 &= (E_i \Lambda_i E^{-1}_i)^H \Lambda_{T_i} (E_i E_i^H)^{-1} E_i \Lambda_i E^{-1}_i \\
			           	 &= \Lambda_{T_i} (E^{-1}_i)^H \Lambda_i^H E_i^H
			           	  				  (E_i^H)^{-1} E_i^{-1}
			           					   E_i \Lambda_i E^{-1}_i \\
			           	 &= \Lambda_{T_i} (E^{-1}_i)^H \Lambda_i^H \Lambda_i E^{-1}_i \\
			           	 &= \Lambda_{T_i} (E_i E_i^H)^{-1} = T.
		\end{aligned}
		\end{align}
		
		Thus, $T_i$ is symmetrical and positive definite and $S_i' T_i S_i = T_i, i \in \mathbb{N}_1^{\bar{q}}$.
		Further, the block diagonal matrix $T_\mathrm{n}$ with $T_i$ as diagonal blocks is symmetrical and positive definite and $S_\mathrm{n}' T_\mathrm{n} S_\mathrm{n} = T_\mathrm{n}$.
		
		Thus, $T=diag(T_\mathrm{p},T_\mathrm{n})$ satisfies (A7).
	\end{proof}
	
	The non-periodic property of reference also brings difficulties in determining terminal constraint $\mathcal{Z}_f$,
	which is concluded in the following lemma.
	
		
		\textbf{Lemma 6:}
		When reference $r(t)$ is non-periodic, $\mathcal{O}_\infty$ corresponding to system (\ref{eqn:augmented_system}) under constraint (\ref{eqn:constraint_augmented_system}) cannot be determined in finite steps.
	
	\begin{proof}
		This is proved by contradiction. 
		
		Suppose $\mathcal{O}_\infty$ is finitely determined as a polytope with a finite number of linear constraints. 
		Since $\mathcal{Z}$ is bounded, $\mathcal{O}_\infty \subseteq \mathcal{Z}$ is bounded.
		Thus, $\mathcal{O}_\infty$ is a convex hull with a finite number of vertexes.
		Further,   $\mathcal{R}_f$ is a convex hull with a finite number of vertexes, as well.
		
		Notate $r^* = \mathrm{max}_{r'T_\mathrm{n} r}  \left\{ r | r \in \mathcal{R}_f \right\}$.
		Ellipsoid $\mathcal{E}_{T}(r^*)$ intersects with $\mathcal{O}_\infty$ at only vertexes. 
		As shown in III-C, $ r \in \mathcal{R}_f$ implies $S^k r^* \in  \mathcal{R}_f, k \in \mathbb{N}^+$.
		Meanwhile, 	since $S' T S = T$, we have $S^k r^* \in  \mathcal{E}_{T}(r^*)$.
		Thus, $S^k r^* \in \mathcal{R}_f \bigcap \mathcal{E}_{T}(r^*)$ are all vertexes of $\mathcal{R}_f$.
		Since $	S^{k} \neq I, \forall k\in\mathbb{N}$, $\mathcal{R}_f$ has an infinite number of vertexes, 
			which is contradicted to the statement that $\mathcal{R}_f$ is a convex hull with a finite number of vertexes.
	\end{proof}
	
	As $\mathcal{O}_\infty$ cannot be finitely determined when reference is non-periodic, 
	$\mathcal{O}_\infty$ cannot be utilized as the terminal constraint.
	In this case, additional steps should be conducted to obtain $\mathcal{Z}_f$. 
			
	The augmented system of (\ref{eqn:augmented_system}) and (\ref{eqn:constraint_augmented_system}) can be rewritten as
	\begin{align}
		\begin{bmatrix} \bar{x}(t+1) \\ \alpha(t+1) \end{bmatrix} &= 
		\begin{bmatrix} A_\mathrm{cl} & 0   \\  0 & S_\mathrm{p}\end{bmatrix}
		\begin{bmatrix} \bar{x}(t) \\ \alpha(t)   \end{bmatrix}		\label{eqn:xalpha}
		\\ 
		\begin{bmatrix} I & \Pi_\mathrm{p}  \\  K & \Gamma_\mathrm{p}  \end{bmatrix}
		\begin{bmatrix} \bar{x}(t)  \\ \alpha(t) \end{bmatrix} &\in \mathcal{Z} 
			\ominus \begin{bmatrix} \Pi_\mathrm{n} \\ \Gamma_\mathrm{n} \end{bmatrix} \beta(t), 	\label{eqn:constriant_xalpha}
		\\
		\beta(t+1) &= S_\mathrm{n} \beta(t),
	\end{align}
	
	where $\Pi = \begin{bmatrix} \Pi_\mathrm{p} & \Pi_\mathrm{n}  \end{bmatrix}, \Gamma = \begin{bmatrix} \Gamma_\mathrm{p} & \Gamma_\mathrm{n} \end{bmatrix}$.
		
	Given $\bar{r}(0) = \begin{bmatrix}	\alpha'(0) & \beta'(0)	\end{bmatrix}'$, 
	since $S_\mathrm{n}' T_\mathrm{n} S_\mathrm{n} = T_\mathrm{n}$,
	we have $\beta(t) \in \mathcal{E}_{T_\mathrm{n}} \big( \beta(0) \big)$ and $\mathcal{E}_{T_\mathrm{n}} \big( \beta(0) \big) = \mathcal{E}_{T_\mathrm{n}} \big( \beta(t)\big), \forall t \in \mathbb{N}_0^+$. 
	To guarantee (A8), the set of non-periodic part of feasible reference is given as
	\begin{align}	\label{eqn:dafwe5}
		\mathcal{R}_f^\mathrm{n} = \left\{ \beta_0 | 
			\mathcal{Z} \ominus \begin{bmatrix} \Pi_\mathrm{n}' & \Gamma_\mathrm{n}' \end{bmatrix}' \beta \neq \emptyset, 
			\forall \beta \in \mathcal{E}_{T_\mathrm{n}}(\beta_0)	\right\}.
	\end{align}
	Clearly, $\begin{bmatrix} \Pi_\mathrm{n}' & \Gamma_\mathrm{n}' \end{bmatrix}' \mathcal{R}_f^\mathrm{n} \subseteq \mathcal{Z}$.
	Moreover, $\mathcal{R}_f^\mathrm{n}$ is an ellipsoid and there exists $\Upsilon \geq 0$ such that $\mathcal{R}_f^\mathrm{n} = \left\{ \beta | \beta' T_\mathrm{n} \beta \leq \Upsilon^2 \right\}$. 
	Since $S_\mathrm{n}' T_\mathrm{n} S_\mathrm{n} = T_\mathrm{n}$, we have $S_\mathrm{n} \mathcal{R}_f^\mathrm{n} = \mathcal{R}_f^\mathrm{n}$.

	Consider $\| \beta \|_{T_\mathrm{n}} \leq \Upsilon$.
	Then, $\mathcal{E}_{T_\mathrm{n}}(\beta) \subseteq \mathcal{R}_f^\mathrm{n}$ and 
	$\mathcal{E}_{T_\mathrm{n}}(\beta) = \dfrac{\| \beta \|_{T_\mathrm{n}}}{\Upsilon} \mathcal{R}_f^\mathrm{n}$.
	Further,
	\begin{align}
	\begin{aligned}
					\mathcal{Z} \ominus \begin{bmatrix} \Pi_\mathrm{n}  \\ \Gamma_\mathrm{n}  \end{bmatrix}  \mathcal{E}_{T_\mathrm{n}}(\beta) 
		= 			\mathcal{Z} \ominus \begin{bmatrix} \Pi_\mathrm{n}  \\ \Gamma_\mathrm{n}  \end{bmatrix}  &\frac{\| \beta \|_{T_\mathrm{n}}}{\Upsilon} \mathcal{R}_f^\mathrm{n} \\
		\supseteq   \mathcal{Z} \ominus  \frac{\| \beta \|_{T_\mathrm{n}}}{\Upsilon} \mathcal{Z}
		= 			 \Big( 1-&\frac{\| \beta \|_{T_\mathrm{n}}}{\Upsilon} \Big) \mathcal{Z}.
	\end{aligned}
	\end{align}
	
	Notate $f(\| \beta \|_{T_\mathrm{n}}) = 1-{\| \beta \|_{T_\mathrm{n}}}/{\Upsilon}$. 
	Then, for $\beta \in  \mathcal{R}_f^\mathrm{n}$, we have $0 \leq f(\| \beta \|_{T_\mathrm{n}}) \leq 1$ and 
	\begin{align}
		f(\| \beta \|_{T_\mathrm{n}})\mathcal{Z} \subseteq \mathcal{Z} \ominus \begin{bmatrix} \Pi_\mathrm{n}' & \Gamma_\mathrm{n}' \end{bmatrix}' \beta  
	\end{align}
	
	According to Lemma 4, the maximal output admissible set $\mathcal{O}^\mathrm{p}_\infty$ corresponding to system (\ref{eqn:xalpha}) under constraint 
	$\begin{bmatrix} I & \Pi_\mathrm{p}  \\  K & \Gamma_\mathrm{p}  \end{bmatrix}
	\begin{bmatrix} \bar{x}(t)  \\ \alpha(t) \end{bmatrix} \in \mathcal{Z}$
	can be determined in finite steps.
	As shown in \cite{gilbert1991linear}, 
	the maximal output admissible set corresponding to system (\ref{eqn:xalpha}) under constraint 
	$\begin{bmatrix} I & \Pi_\mathrm{p}  \\  K & \Gamma_\mathrm{p}  \end{bmatrix}
	\begin{bmatrix} \bar{x}(t)  \\ \alpha(t) \end{bmatrix} \in {\mu} \mathcal{Z}, \mu\geq 0$
	is ${\mu} \mathcal{O}^\mathrm{p}_\infty$.
	Define $\mathcal{R}^\mathrm{p}_\infty = \{ \alpha | \begin{bmatrix} 0 & \alpha'	\end{bmatrix}' \in \mathcal{O}^\mathrm{p}_\infty\}$.
	Then, $\mathcal{O}^\mathrm{p}_\infty$ and $\mathcal{R}^\mathrm{p}_\infty$ can be determined offline.
	
	The terminal constraint and properties of the corresponding $\mathcal{R}_f$ and $\mathcal{R}_f^\mathrm{m}$ are given in the following lemma.
	
		
		\textbf{Lemma 7:}
		(i)
		If (A2) and (A3) hold and reference is non-periodic, 
		$\mathcal{Z}_f = f(\| \beta \|_{T_\mathrm{n}}) \mathcal{O}^\mathrm{p}_\infty \times \mathcal{R}_f^\mathrm{n}$ satisfies (A8);
		(ii) $\mathcal{R}_f = \mathcal{R}_f^\mathrm{m}$ where
		\begin{align}
			\mathcal{R}_f := \left\{ r =	\begin{bmatrix}	\alpha' & \beta'	\end{bmatrix} ' |
			\alpha \in f(\| \beta \|_{T_\mathrm{n}}) \mathcal{R}^\mathrm{p}_\infty, \beta \in  \mathcal{R}_f^\mathrm{n}
			\right\}.
		\end{align}
	
	\begin{proof}
		For all $ \begin{bmatrix}	\bar{x}' & {r}'	\end{bmatrix} '  \in f(\| \beta \|_{T_\mathrm{n}}) \mathcal{O}^\mathrm{p}_\infty \times\mathcal{R}_f^\mathrm{n}$,
		we have $\beta \in \mathcal{R}_f^\mathrm{n}$ and $ f(\| \beta \|_{T_\mathrm{n}}) \geq 0$.
		Further, according to the property of $\mathcal{O}^\mathrm{p}_\infty$,
		$\begin{bmatrix} I & \Pi_\mathrm{p}  \\  K & \Gamma_\mathrm{p}  \end{bmatrix}
		\begin{bmatrix} \bar{x}  \\ \alpha \end{bmatrix} 
		\in f(\| \beta \|_{T_\mathrm{n}}) \mathcal{Z} 
		\subseteq \mathcal{Z} 
		\ominus \begin{bmatrix} \Pi_\mathrm{n}' & \Gamma_\mathrm{n}' \end{bmatrix}' \beta$. Thus, 
		\begin{align}
			\begin{bmatrix} I & \Pi  \\  K & \Gamma  \end{bmatrix}
			 \begin{bmatrix}	\bar{x} \\ r	\end{bmatrix} = 			
			\begin{bmatrix} I & \Pi_\mathrm{p} & \Pi_\mathrm{n} \\  K & \Gamma_\mathrm{p} & \Gamma_\mathrm{n} \end{bmatrix}
			\begin{bmatrix} \bar{x}  \\ \alpha \\ \beta \end{bmatrix} \in \mathcal{Z}.
		\end{align}
		
		Since $\beta \in \mathcal{R}_f^\mathrm{n}$, we have $S_\mathrm{n} \beta \in S_\mathrm{n} \mathcal{R}_f^\mathrm{n} = \mathcal{R}_f^\mathrm{n}$.
		Meanwhile, according to the property of $\mathcal{O}^\mathrm{p}_\infty$, we have
		$\begin{bmatrix}	(A_\mathrm{cl}\bar{x})' & (S_\mathrm{p} r)'	\end{bmatrix}' \in f(\| \beta \|_{T_\mathrm{n}}) \mathcal{O}^\mathrm{p}_\infty$.
		Then,
		\begin{align}
			\begin{bmatrix} A_\mathrm{cl} & 0  \\  0 & S  \end{bmatrix}
			\begin{bmatrix} \bar{x} \\ r   \end{bmatrix}
			\in f(\| \beta \|_{T_\mathrm{n}}) \mathcal{O}^\mathrm{p}_\infty \times \mathcal{R}_f^\mathrm{n}.
		\end{align}
		
		Thus, (i) holds.
		
		The proof of (ii) is similar to the proof of Lemma 4 and thus, is omitted here.
	\end{proof}
	
		
		\textbf{Remark 4:}
		In practice, we suggest to  replace $\beta$ by $\beta^\dagger = sat(\frac{\| \beta \|_{T_\mathrm{n}}}{\Upsilon}) \frac{\Upsilon}{\| \beta \|_{T_\mathrm{n}}} \beta$
			before conducting optimization problem (\ref{eqn:QP}).
		By doing so, $\beta \in  \mathcal{R}_f^\mathrm{n}$ is ensured.
		Then, terminal constraint $\mathcal{Z}_f$ can be replaced by $\mathcal{Z}_f^\dagger = f(\| \beta^\dagger \|_{T_\mathrm{n}}) \mathcal{O}^\mathrm{p}_\infty \times \mathbb{R}^{q_\mathrm{n}}$ so that only linear constraints are involved.
		In this case, $\begin{bmatrix} \bar{x}'(N|t)  &  \bar{r}'(N|t) \end{bmatrix}' \in \mathcal{Z}_f^\dagger$ implies $\begin{bmatrix} \bar{x}'(N|t)  &  \bar{r}'(N|t) \end{bmatrix}' \in \mathcal{Z}_f$.
		Thus, properties of the proposed MPC controller given in Theorem 1 still holds.

\section{Examples}
	The system model used in this example comes from \cite{rakovic2023model}.
	The components of state are horizontal position, vertical position, horizontal velocity, vertical velocity, pitch rate, and pitch angle. 
	The components of control input $u$ are ``collective" and ``longitudinal cyclic" pitch control.
	The continuous-time model in \cite{rakovic2023model} is discretized with period $T_\Delta = 0.5$. 
	Matrices $A,B,C$ and $Q_e$ are given as
	\begin{align*}
		\begin{aligned}
			A =\begin{bmatrix} 1 & 0 & 0.4954 & 0.0026 & -0.0069 & -0.0596 \\
				0 & 1 & 0.0042 & 0.3896 & -0.0688 & -0.4395 \\
				0 & 0 & 0.9813 & 0.0083 & -0.0454 & -0.2459 \\
				0 & 0 & 0.0117 & 0.5813 & -0.3898 & -1.6662 \\
				0 & 0 & 0.0457 & 0.1274 & ~~0.8230 &  ~~0.4803 \\
				0 & 0 & 0.0117 & 0.0358 & ~~0.4433 &  ~~1.1361 \end{bmatrix}, 
				\\
			B =\begin{bmatrix} ~~0.0609 &  ~~0.0148 \\
				~~0.4255 &   -0.8451 \\
				~~0.2664 &  ~~0.0365 \\
				~~1.7629 &   -3.2664 \\
				-2.3452  &  ~~1.7209 \\
				-0.6083  &  ~~0.4660 \end{bmatrix},
			C  =\begin{bmatrix}1&0\\0&1\\0&0\\0&0\\0&0\\0&0\end{bmatrix}',
			Q^e=\begin{bmatrix}1&0\\0&1\\1&0\\0&1\\1&0\\0&1\end{bmatrix}'.
		\end{aligned}	
	\end{align*}
	
	$S$ is given as $S  = diag(S_1,S_2,S_3,S_4), S_1  = I, S_i = e^{S_i^\mathrm{c} T_\Delta}, i= 2,3,4$, 
	where $S_1^\mathrm{c} = -\frac{ \pi}{24} \begin{bmatrix}	0 ~~ 1; ~~ -1 ~~ 0	\end{bmatrix},
	       S_2^\mathrm{c} = -\frac{ \pi}{4}  \begin{bmatrix}	-s_0 ~~ 1; ~   -1 ~~ s_0 \end{bmatrix}, 
	       S_3^\mathrm{c} = -\frac{   1}{20} \begin{bmatrix}	0 ~~ 11/10 ; ~~ -10/11 ~~ 0	\end{bmatrix}$, 
	and $s_0 = \cos(-0.45 \pi)$.
	In this example, $S_\mathrm{p} =$ $ diag( S_1,S_2,S_3 )$, $S_\mathrm{n} = S_4$, and $S_\mathrm{p}^{96} = I$.

	Constraints on system (\ref{eqn:actualSystem}) is given as	$\mathcal{Z} = \big\{\begin{bmatrix} x' & u'	\end{bmatrix}' | $
	$x \in \mathcal{X}_1 \times \mathcal{X}_2, 	u \in \mathcal{U}	\big \}$, 
	where 
	$\mathcal{X}_1  = \left\{ x \in  \mathbb{R}^2 | \left \| x_i \right \| _\infty  \leq 50  \right\}$,  
	$\mathcal{X}_2  = \left\{ x \in  \mathbb{R}^4 | \left \| x_i \right \| _\infty  \leq 1  \right\}$,  
	$\mathcal{U}  = \left\{ u \in  \mathbb{R}^2 | \left \| u   \right \| _\infty  \leq 0.2  \right\}.$
	
	The feedback gain is given as $K   = \begin{bmatrix} K_1 & K_2\end{bmatrix}$, where
	\begin{align*}
		K_1  &= \begin{bmatrix} -0.3435 & 0.1540 & -0.9801 \\ 
			-0.1795 & 0.2915 & -0.4789 \end{bmatrix}, \\
		K_2 &= \begin{bmatrix}  ~~0.1734 & 0.4237 &  ~~0.6930 \\ 
			~~0.2935 & 0.0743 & -0.1997 \end{bmatrix}.
	\end{align*}
	
	$\Pi,\Gamma$ and $L$ are determined correspondingly.
	
	Given $Q = I$, $P$ is determined by solving the discrete-time Lyapunov equation. 
	$T$ is chosen as $T = diag(T_\mathrm{p}, T_\mathrm{n})$ 
	where $T_\mathrm{p} = \sum_{i=1}^{96} {S_\mathrm{p}^i}' T_0 S_\mathrm{p}^i$ with $T_0 = I$ 
	and $T_\mathrm{n} =  \Lambda_{T_\mathrm{n}} (E E^H)^{-1}$ with $\Lambda_{T_\mathrm{n}} = 100I$.
	The prediction horizon is $N=10$.
	
	The reference contains two parts.
	The first part begins with $r(0) = \begin{bmatrix}	    -2 & -2 & -1 & -1.5 & -0.1 & -0.15 & 0 & 0	\end{bmatrix}'$.
	At $t = 251$, reference switches to the second part with $	r(251) = \begin{bmatrix}	2 & 6 & 0 & 2 & 0 & 0 & 0 & 2 	\end{bmatrix}'$.
	The first part of reference is periodic and the second part is non-periodic.
	
\begin{figure}
		\centerline{\includegraphics[width=\columnwidth]{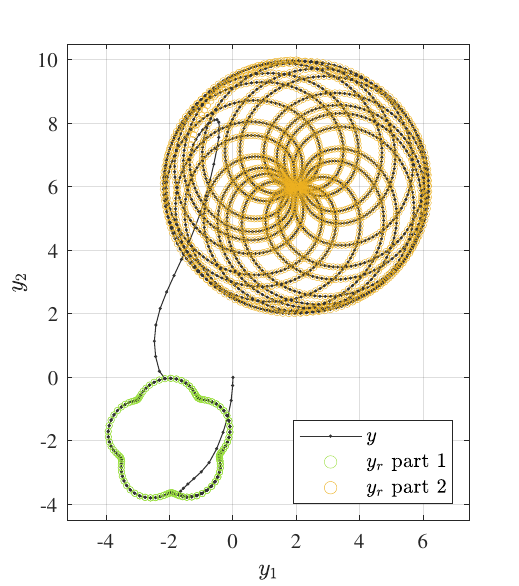}}
		\caption{Trajectory of output $y$ and desired output $y_r$.}
		\label{fig:yyr}
	\end{figure}

\begin{figure}
		\centerline{\includegraphics[width=\columnwidth]{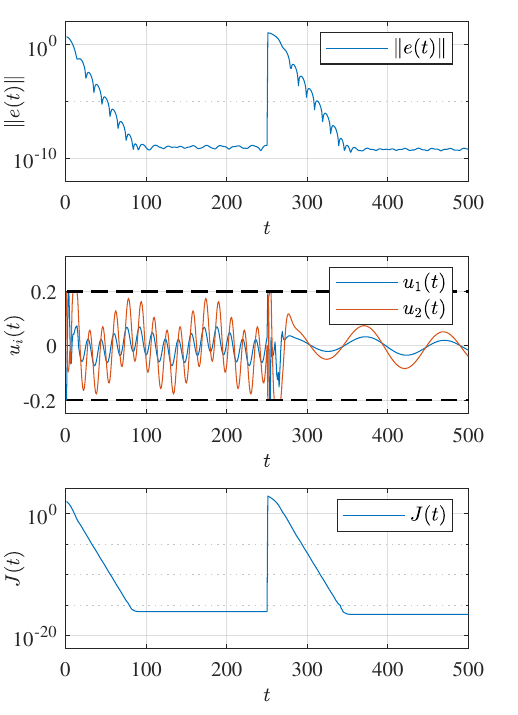}}
		\caption{Trajectories of the norm of tracking error $\|e(t)\|$, control input $u(t)$, and cost $J(t)$.}
		\label{fig:euj}
	\end{figure}

	With $x(0)=0$, trajectories of the output and the desired output from $t=0$ to $t=2500$ are shown in Fig. \ref{fig:yyr}. 
	Fig. \ref{fig:euj}. illustrates trajectories of the norm of tracking error, control inputs, and cost from $t=0$ to $t=500$.
	As shown in Fig. \ref{fig:euj}, tracking error converges to $0$.
	During the convergence, constraints on control input are satisfied.
	Meanwhile, cost increases only at the switching step $t=251$ and is non-increasing during the rest time.
	It should be noted that if we specify $\bar{r}(t) = r(t)$, optimization problem (\ref{eqn:QP}) becomes infeasible at $t=251$, 
		which illustrates the necessity of artificial reference in dealing with sudden changes of reference.
	
	
	\textbf{Remark 5:}
	The period of the first part is $96$.
	With the proposed method, the number of additional decision variables is $8$.
	In contrast, the MPC method proposed in \cite{limon2015mpc} introduces $192$ more decision variables.
	This implies that the proposed method has a lower computational burden.
	Moreover, the second part of reference is non-periodic, which is not considered in \cite{limon2015mpc}.
	Thus, it is more efficient to use the MPC method proposed in this article to track references with specific dynamics.
	
	\section{Conclusion}
	In this article, an MPC controller is proposed for constrained linear systems to track bounded references with arbitrary dynamics.
	The proposed method includes artificial reference as decision variable so that it can deal with sudden changes of reference.
	The dynamics of reference imposes difficulties in determining terminal constraint and cost function.
	When reference is periodic, a suitable weight matrix is determined, and terminal constraint is chosen as the maximal output admissible set which is proved to be finitely determined.
	When reference is non-periodic, a suitable weight matrix can be found, as well, but the maximal output admissible set cannot be  finitely determined.
	In this case, periodic and non-periodic parts of reference are analyzed separately to determine a finitely-determined terminal constraint.
	Future research interest includes developing robust MPC controller for reference tracking and investigating the potential for tracking some types of unbounded references.

	\section*{References}
	\bibliographystyle{IEEEtran}
	\bibliography{biblio_MPC_tracking_TAC}
	
\end{document}